\documentclass[12pt]{article}
\usepackage{subeqn}
\usepackage{fullpage}

\newcommand\be{\begin{equation}}
\newcommand\ee{\end{equation}}
\newcommand\bea{\begin{eqnarray}}
\newcommand\eea{\end{eqnarray}}
\newcommand\bean{\begin{eqnarray*}}
\newcommand\eean{\end{eqnarray*}}
\newcommand\nn{\nonumber}

\newcommand\bdm{\begin{displaymath}}
\newcommand\edm{\end{displaymath}}

\def\pmb#1{\setbox0=\hbox{#1}%
  \kern-.025em\copy0\kern-\wd0
   \kern.05em\copy0\kern-\wd0
   \kern-.025em\raise.0433em\box0 }
\def\btimes{\pmb{${\bf \times}$}}  

\def\rd{\mathrm d}
\def\rDD{\mathrm D}

\begin{document}

\title{\large {\bf  A generalization of the F\'enyes-Nelson stochastic \\model of quantum mechanics\thanks{{\em Letters in Mathematical Physics\/} {\bf 3}  (1979) 271--277. \copyright  1979 {\em D.\ Reidel Publishing Company.}}}}

\author{{\normalsize Mark Davidson}\thanks{
Current Address: Spectel Research Corporation, 807 Rorke Way, Palo Alto, CA   94303
\newline  Email:  mdavid@spectelresearch.com, Web: www.spectelresearch.com}\\
\normalsize{\em }}

\date{}

\maketitle
\begin{abstract} It is shown that the stochastic model of F\'enyes and Nelson can be generalized in such a way that the diffusion constant of the Markov theory becomes a free parameter.  This extra freedom allows one to identify quantum mechanics with a class of Markov processes with diffusion constants varying from 0 to $\infty$.
\end{abstract}

\section{Introduction}

In 1952 Imre F\'enyes discovered that Schr\"odinger's equation could be interpreted as a diffusion equation for a continuous Markov process \cite{fenyes1}.  This result occured in the wake of the great debates over the interpretation of quantum mechanics of the 1920's and 30's.  It crystallized earlier and decidedly stochastic concepts found in the works of Schr\"odinger, Bopp, De Broglie, Einstein, and many others, who questioned the completeness of Bohr's complementarity interpretation of quantum mechanics.  In the 1960's, Edward Nelson rederived the F\'enyes result  and expanded it  considerably, using the rigorous and powerful tools of modern probability theory \cite{nelson1,nelson2}.  This elegant and important work influenced successive developments in the field.  Following Nelson's work, de la Pena-Auerbach made numerous significant contributions to the field \cite{auerbach1,auerbach2}.  The extraordinary book of Max Jammer \cite{jammer} has a good review of the subject, as does an article by Claverie and Diner \cite{claverie}.  More recent contributions include the work of Dankel \cite{dankel} on spin, Guerra \cite{guerra} and Albeverio \cite{albeverio} on relativistic fields, and Shucker \cite{shucker} on the asymptotic behavior of sample trajectories.  Moyal mechanics \cite{moyal} is the main competitor to the F\'enyes-Nelson model as a stochastic basis for quantum mechanics.  A Langevin approach to the problem of an electron in stochastic electrodynamics \cite{auerbach3} supports the Moyal picture, and both models deserve serious study.

A peculiarity of the process which Nelson and F\'enyes studied is that the acceleration of the diffusing particle cannot be uniquely defined.  This allows many possibilities when constructing a dynamical theory.  Nelson showed that a particular dynamical assumption, equating the `mean acceleration' to the external force, leads to Schr\"odinger's equation as a solution to the Markov diffusion problem.  The point of this paper is to show that this procedure is not unique, and to illustrate a whole class of possible dynamics, all of which yield Schr\"odinger's equation.  The result is that the diffusion constant of the Markov theory is not determined, but  rather can take on any positive value.  Moyal mechanics can consequently be viewed as a limit of F\'enyes-Nelson type models.  This was reported earlier using a new mathematical formalism \cite{davidson}.  Here it is proved using Nelson's formalism.

\section{Generalized stochastic dynamics}
We begin the discussion with a theorem about Schr\"odinger's equation:
\be
\left[ -\frac{\hbar^2}{2m} \Delta +V\right] \exp (R +iS ) =i\hbar \frac{\partial}{\partial t} \, \exp (R +iS).
\label{1}
\ee
Consider the following equation, where $z$ is a real parameter:
\bea
\left[ -\frac{(z\hbar)^2}{2m} \Delta \right.&+&\left.\left( V+ \frac{\hbar^2}{2m} \,(z^2 -1) \frac{\Delta \sqrt\rho}{\sqrt \rho}\right) \right] \exp (R+iS/z)\nn\\
&& =i(z\hbar) \,\frac{\partial}{\partial t} \,\exp (R+iS/z)
\label{2}
\eea
where
\be
\rho =\exp (2R)
\label{3}
\ee
is the probability density for the particle.

\noindent {\sc Theorem 1}. {\em  Let $R$ and $S$ be bounded, continuous, real functions of $\vec x$ and $t$ with first and second space derivatives and first time derivatives in a region $\Omega$ in $(\vec x, t)$ space.  Let $z\not= 0$ in eqn.\ {\rm (\ref{2}) } above.  Then eqns.\ {\rm (\ref{1})} and {\rm (\ref{2})} are equivalent for $(\vec x , t)$ in $\Omega$.}

\noindent{\em Proof}.  Let $(\vec x, t)$ be in $\Omega$.  Divide eqn.\ (\ref{1}) by $\exp (R+iS)$ and divide eqn.\ (\ref{2}) by $\exp (R+iS/z)$.

\noindent Equation (\ref{1}) becomes:
\bea
-\frac{\hbar^2}{2m} \, \biggl\{\biggr. \Delta R &+&(\vec \nabla R)^2 -(\vec\nabla S)^2 + i\Delta S+ i2\vec\nabla
R\cdot\vec\nabla S\biggl.\biggr\} +V\nn\\
&&\quad =i\hbar (\dot R+ i\dot S)
\label{4}
\eea
and eqn.\ (\ref{2})   becomes:
\bea
-\frac{(z\hbar)^2}{2m}\biggl\{ \Delta\biggr. R &+& (\vec \nabla R)^2 -\frac{(\vec\nabla S)^2}{ z^2} +i\Delta S/z +i 2\vec\nabla R\cdot \vec\nabla S/z\biggl.\biggr\}\nn\\
&&\quad +V + (z^2 -1) \, \frac{\hbar^2}{2m} \, \frac{\Delta \sqrt \rho}{\sqrt \rho} \,= i(z\hbar ) (\dot R + i \dot S/z).
\label{5}
\eea
The real part of eqn.\ (\ref{4}) is
\be
-\frac{\hbar^2}{2m} \left\{ \Delta R + (\vec\nabla R)^2 - (\vec\nabla S)^2\right\} +V= -\hbar \dot S.
\label{6}
\ee
The real part of eqn.\ (\ref{5}) is
\be
-\frac{(z\hbar)^2}{2m} \left\{ \Delta R +(\vec\nabla R)^2 -\frac{ (\vec\nabla S)^2}{z^2} \right\} +V+(z^2 -1) \frac{\hbar^2}{2m} \, \frac{\Delta \sqrt \rho}{\sqrt \rho} =-\hbar \dot S
\label{7}
\ee
but,
\be
\frac{\Delta \sqrt \rho}{\sqrt \rho} =\Delta R + (\vec\nabla R)^2.
\label{8}
\ee
Equations (\ref{6}) and (\ref{7}) are seen to be identical.  The imaginary part of (\ref{4}) is:
\be
-\frac{\hbar^2}{2m} \left\{ \Delta S+2\vec\nabla R\cdot \vec\nabla  S\right\} =\hbar \dot R.
\label{9}
\ee
The imaginary part of eqn.\ (\ref{5}) is
\be
-\frac{ (z\hbar)^2}{2m} \left\{ \frac{\Delta S}{z} +2\vec \nabla R\cdot \vec\nabla S/z\right\} =z\hbar \dot R.
\label{10}
\ee
Equations (\ref{9}) and (\ref{10}) are also identical, and therefore eqns.\ (\ref{3}) and (\ref{4}) are equivalent.  The theorem follows immediately.

The stochastic process which is used to model quantum mechanics is defined as follows \cite{nelson1,nelson2}:
\be
\rd \vec x =\vec b (\vec x, t) \rd t +\rd \vec W.
\label{11}
\ee
In this Langevin equation, $\vec W$ is a Wiener process.  The solution to (\ref{11}) is a continuous Markov process provided certain conditions on $\vec b$ are met.  The sample trajectories are nowhere differentiable functions of time almost surely, and as a consequence, velocities, accelerations, and higher time derivatives cannot be introduced unambiguously. One way of proceeding to a dynamical theory is to introduce operators on a Hilbert space corresponding to velocity, acceleration, etc.\ \cite{davidson}.  Another way is to introduce forward and backward time derivatives $D$ and $D_*$, respectively.  Following Nelson (Ref.\ 2, p.\ 104), we may define
\bean
\rDD f(x,t) &=&\lim\limits_{h\to 0_+} \, E\left( \frac{f(x(t+h), t+h)-f(x(t),t)}{h}\biggl.\biggr| \, x(t) =x\right),\nn\\
\rDD_* f(x,t) &=&\lim\limits_{h\to 0_+}  E\left( \frac{f(x(t), t) -f(x(t-h), t-h)}{h} \biggl.\biggr| x(t)=x\right),
\eean
and then it may be shown:
\bea
\rDD &=& \frac{\partial}{\partial t} +\vec b \cdot \vec \nabla + \nu \Delta,\label{12}\\
\rDD_* &=&\frac{\partial}{\partial t} +\vec b_* \cdot \vec\nabla -\nu \Delta,
\label{13}
\eea
where $\vec b_*$ is defined by:
\be
(\vec b -\vec b_*)/2 =2\nu \vec\nabla R=\nu \vec\nabla \ln (\rho).
\label{14}
\ee
Clearly the fact that $\rDD$ and $\rDD_*$ are different operators is intimately connected to the fact that the sample trajectories are not differentiable.  If the sample trajectories were smooth, then $\rDD$ and $\rDD_*$ would have to be the same.  Thus it is impossible to ascribe an instantaneous acceleration to the  diffusing particle.  As a proxy, one may consider quadratic expressions in $\rDD$ and $\rDD_*$ operating on $\vec x$. For example, Nelson chose the following dynamical assumption:
\be
\frac{m}{2} \left[\rDD \rDD_* +\rDD_* \rDD\right] \vec x =-\vec\nabla V
\label{15}
\ee
where $V$ is the external potential. He showed (Ref.\ 2, Chap.\ 15) that (\ref{15}) leads to the equation
\be
\left[ -\frac{(2m\nu)^2}{2m} \Delta +V\right] \exp (R+iS_N) =i(2m\nu) \frac{\partial}{\partial t} \,\exp (R+iS_N)
\label{16}
\ee
provided that
\be
\vec\nabla {\btimes} (\vec b +\vec b_*)=0,
\label{17}
\ee
and where
\be
\vec b =2\nu \vec \nabla (R+S_N)
\label{18}
\ee
defines $S_N$ up to an arbitrary additive function of $t$.  If $\nu =\hbar /2m$ then (\ref{16}) is just Schr\"odinger's equation.

There are other possibilities besides eqn\ (\ref{15}).  Consider the dynamical equation:
\be
(m/2) \left[\rDD \rDD_* +\rDD_* \rDD\right] \vec x +m (\beta/8) (\rDD -\rDD_*)^2 \vec x =\vec \nabla V
\label{19}
\ee
where $\beta$ is an arbitrary real constant.  In the limit $\rDD =\rDD_*$, $\nu\to 0$, eqns.\ (\ref{15}) and (\ref{19}) become equivalent.  Since $D$ and $D_*$ appear symmetrically in (\ref{19}) there is no preferred direction in time implicit in this equation.  The extra term in (\ref{19}) may be reexpressed using (\ref{12}) and (\ref{13}):
\be
m(\beta /8) (\rDD -\rDD_*)^2 \vec x =\vec \nabla \left[ m\beta\nu^2 \frac{\Delta \sqrt \rho}{\sqrt \rho}\right],
\label{20}
\ee
so that (\ref{19}) becomes:
\be
\frac{m}{2} \left[\rDD \rDD_* +\rDD_*\rDD \right] \vec x =-\vec\nabla \left( V+m\beta\nu^2 \frac{\Delta\sqrt \rho}{\sqrt \rho}\right).
\label{21}
\ee
This has the same form as eqn.\ (\ref{15}) except that there is an extra term in the potential. Using the result (\ref{16}), eqn.\ (\ref{21}) yields:
\be
\left[ -\frac{(2m\nu)^2}{2m} \Delta +V +\frac{\beta}{2} \,\frac{(2m\nu)^2}{2m} \, \frac{\Delta \sqrt \rho}{\sqrt \rho} \right] \exp (R+iS_N) =i (2m\nu)\,\frac{\partial}{\partial t} \,\exp (R+iS_N).
\label{22}
\ee
Now compare this equation with eqn.\ (\ref{2}) and Theorem 1.  If $z$ satisfies:
\be
z\hbar =2m\nu
\label{23}
\ee
and also
\be
\frac{\hbar^2}{2m} (z^2 -1) =\frac{\beta}{2} \, \frac{(2m\nu)^2}{2m}
\label{24}
\ee
or equivalently
\be
z=1 /\sqrt{1-\beta/2}
\label{25}
\ee
then by Theorem 1, eqn\ (\ref{22}) is equivalent to:
\be
\left[ -\frac{\hbar^2}{2m} \Delta +V\right] \exp (R+izS_N) = i\hbar \frac{\partial}{\partial t} \, \exp (R+izS_N).
\label{26}
\ee
Defining:
\be
\psi =\exp (R+izS_N)
\label{27}
\ee
then (\ref{26}) is just Schr\"odinger's equation.  The parameter $\nu$ can be chosen to be any positive constant. If $\beta=0$, then $\nu =\hbar /2m$ which is Nelson's result.  In Ref.\ 14, a similar result was presented ($\nu$ in (\ref{14}) differs from  Nelson's definition, used here, by a factor of 2).

The fact that $zS_N$ appears in $\psi$ rather than simply $S_N$ does not appear to cause conceptual difficulties. For example, the expected value of the momentum is:
\be
\int \rd^3 x \psi^* (-i\hbar \vec\nabla )\psi =\int \rd^3 x\rho z\hbar \vec \nabla S_N =\int \rd^3 x\rho 2 m\nu \vec \nabla S_N
\label{28}
\ee
which is consistent with the interpretation of $2\nu \vec\nabla S_N$ as a mean flow velocity.  Thus the identification of $-i\hbar \vec \nabla$ as a momentum operator in the Schr\"odinger representation is possible, regardless of the value of $\nu$.  The initial conditions on $\psi$ at $t=0$, say, would be the same regardless of $\nu$.

Although in general difficult, it is in principle possible to calculate the Markov transition function given $\psi$ at $t=0$ and given $\nu$.  The problem becomes much easier for stationary states.  The transition function, expressed as a density, must satisfy the forward equation:
\be
\frac{\partial}{\partial t}  P\,(x, t; y,s) +\vec\nabla_x \cdot \vec b (x,t) \, P (x,t ; y, s) -\nu \Delta_x \,P (x,t ; y,s) =0.
\label{29}
\ee
For stationary states, this becomes:
\be
\frac{\partial}{\partial t} \, P_{t-s} (x,y) +\vec \nabla_x \cdot \nu (\vec\nabla \ln (\rho)) P_{t-s} (x,y) -\nu \Delta_x P_{t-s} (x,y)=0
\label{30}
\ee
with initial conditions following from continuity:
\be
P_0 (x,y) =\delta^3 (\vec x -\vec y).
\label{31}
\ee
The solutions to (\ref{30}) will clearly depend on $\nu$ except in the limit $t\to \infty$ where:
\be
P_\infty (x,y) =\rho (x).
\label{32}
\ee
If Nelson's argument \cite{nelson1}, that Schr\"odinger's equation contains all of the experimental content of quantum mechanics, is accepted, then the fact that the Markov transition functions are different will not lead to experimentally measurable differences.  In fact, since the Markov transition function depends on $R$ and $S_N$, and since any attempt to measure the transition function will inevitably change $R$ or $S_N$, it is impossible to determine the transition function from experiment.  It must be concluded that if the process which Nelson or F\'enyes proposed can yield a consistent and satisfactory model of quantum mechanics, then any of the infinite class of models presented here are equally satisfactory.

These results may be generalized to systems with any number of degrees of freedom, including many particle systems and fields.  They may also be generalized to include magnetic forces.  The result remains true.  Any value of $\nu$ greater than 0 is suitable for a stochastic model of the quantized system.

\section{Conclusion}

The diffusion parameter is not uniquely determined when constructing Markov models of quantum mechanics.  The stochastic interpretation of quantum mechanics must be amended by identifying quantum mechanics with a whole family of Markov processes which have different diffusion constants $\nu$.  If $\nu$ is very small, then the Markov theory is approximately a phase space description in the sense that $\rDD$ and $\rDD_*$ are approximately equal.  In this way phase space models of quantum mechanics, such as Moyal mechanics \cite{moyal,auerbach3}, can be included within this interpretation as a limiting case $(\nu\to 0)$ of the Markov description.

Whether these results hinder or help the stochastic interpretation of quantum mechanics is a matter of taste.  On the one hand, it would have been nice to associate quantum mechanics with a unique stochastic process.  On the other hand, the extra freedom presented here may make it easier to construct relativistically invariant theories, and to extend the stochastic interpretation to more general dynamical systems.

\noindent ({\em Received  February 15, 1979})

\end{document}